# Structure formation and thermal conduction in polymer-based composites obtained by fused filament fabrication

Anton Smirnov,[a] Nestor Washington Solis Pinargote,[a] Roman Khmyrov,[b] Nikolai Babushkin,[a] Mikhail Gridnev,[a] Ekaterina Kuznetsova,[a] Andrey Gusarov[b*]

[a] *Laboratory of 3D Structural and Functional Engineering, Moscow State University of Technology STANKIN, Vadkovsky per. 1, Moscow 127055, Russia*

[b] *Laboratory of Innovative Additive Technologies, Moscow State University of Technology STANKIN, Vadkovsky Per. 3a, 127055 Moscow, Russia*

[*] Corresponding author e-mail: av.goussarov@gmail.com

**Abstract.** Fused filament fabrication (FFF) is widely used to obtain polymer-based composites with improved mechanical and thermal conduction properties. The effective properties of the composites are sensitive to its structure including the shape, distribution, and orientation of inclusions in the matrix and FFF-specific defects. The present work aims to study the formation of composite structure and relate it to the measured effective thermal conduction properties. Polymer flow in the hot end and the structure of printed samples are studied by metallography. The effective thermal diffusivity is measured by the laser flash method and analyzed by the Maxwell Garnett theory. Fibers in a polymer-matrix composite visualize polymer flow in the hot end of the FFF printer. In the nozzle, fibers are generally oriented parallel to the flow direction while gas bubbles may locally disturb flow and disorient fibers. Printed composite Carbon fiber/TPU inherits preferential orientation of fibers observed in the filament. Disorientation of fibers relative to the alignment direction increases after printing. In the printed $Al_2O_3$/PA composite, there is a lattice of triangle-based cylindrical pores parallel to the extrusion direction. $Al_2O_3$ particles are of irregular shape and approximately equiaxed. No segregation or agglomeration are observed. Significant anisotropy of thermal diffusivity is observed in the printed samples of 5%vol. Carbon fiber/TPU composite. The measured effective thermal diffusivity of printed $Al_2O_3$/PA composites is essentially isotropic and significantly increases with the volume fraction of inclusions. In the studied FFF-printed composites, the effective thermal diffusivity is proportional to the thermal diffusivity of polymer matrix, which is sensitive to its structure formed in non-equilibrium conditions of rapid quenching from liquid phase.

# 1. Introduction

Fused filament fabrication (FFF) also known as fused deposition modeling (FDM) is a kind of material extrusion additive manufacturing (AM). This is a low cost and powerful AM technique initially developed for a wide variety of thermoplastic polymers [1]. A polymer filament is melted in the hot end of the printing head and extruded through a nozzle to form a piece of added material on the surface of a growing part. The shape of the part is controlled by positioning the head. Typical polymers for FFF are acrylonitrile butadiene styrene (ABS), polylactic acid (PLA), polycarbonate (PC), polyether ether ketone (PEEK), and Nylon [2]. Filled polymers can be used as well. The advantage is obtaining polymer-based composite materials with improved mechanical and functional properties [3]. To improve mechanical properties, reinforcement by carbon, Kevlar, and glass fibers can be applied [4]. Wear resistance of Nylon 6 was improved by adding aluminum (Al) and alumina ($Al_2O_3$) inclusions [5]. Two-stage processes have been proposed for glass [6], ceramics [7], and metals [8]. In the first FFFF stage, a green part is obtained from a polymer-matrix composite highly loaded with particles of the principal material. The second stage is de-binding and sintering of the green part in a furnace aiming removal of the polymer and consolidation of the principal material.

To obtain conductive parts by FFF, Bin Hamzah *et al.* proposed to add equiaxed carbon inclusions [9] or carbon fibers [10] to the matrix polymer. Wang *et al.* reported a considerable increase of thermal conduction in similar composite materials with $Al_2O_3$ fibers and its dependence on fibers orientation [11]. They concluded that controlling the orientation of fibers is an effective technique to improve thermal conduction. Currently, polymer-based materials with elevated thermal conductivity attract attention of electronic-products developers because they are effective for energy dissipation [12]. Additive manufacturing by FFF printing is a promising technique for manufacturing energy management systems for light-emitting diode devices, electronic packaging, and solar cells [13]. It is useful printing highly thermally conductive but electrically insulating polymer-matrix composites [14].

Since loading of polymers considerably influences their thermal and mechanical properties, this affects thermo-mechanical processes at extrusion defining the productivity of FFF and the quality of obtained material [15]. Heat transfer and flow of softened polymer-based composite in the hot end of the printing head are key physical processes. Understanding these processes is helpful to find optimal printing parameters such as the hot end temperature and the feed rate and to relate them to the composite rheological properties and thermal conduction properties [16]. Shadvar *et al.* [17] numerically simulated coupled heat transfer and non-Newtonian flow in the FFF nozzle and explained the experimentally observed so-called die swell phenomenon. Die swell is the increase of extruded polymer diameter after the exit from the nozzle. Mishra *et al.* [18] implemented Cross-Williams-Landel-Ferry non-Newtonian rheology model and accounted for the density dependence on temperature and pressure. They concluded that insufficient heating of the melt in the nozzle considerably reduces flow rate. To explain experimental data, Ufodike and Nzebuka [19] assumed a non-ideal thermal contact between the liquefier wall and the filament in their numerical model of FFF.

In summary, polymer-matrix composites are widely used for FFF additive manufacturing to improve mechanical and functional properties. In particular, composites with elevated thermal conduction properties are demanded to improve thermal management of electronic devices. The influence of fillers on thermal conduction is, generally, well studied [20]. However, non-ideal structures with non-uniform distribution of inclusions and preferential orientation of fiber inclusions can be formed by FFF. Relations between the specific composite structures and thermal conduction are insufficiently studied. Thermal conduction is also important for the physics of extrusion and should be considered to find optimal FFF process parameters. The physics of extrusion is well studied by numerical simulation but experimental studies are limited. For example, there are known measurements of temperature and flow rate at the outlet of the nozzle [21] and the so-called die swell [17]. However, the experimental approaches to study the flow field in the hot end are necessary to justify the commonly used assumption of laminar flow and to verify vortex flows predicted by simulations [19].

The present work concerns experimental studying the structure of polymer-based composites obtained by FFF and their thermal conduction properties to reveal the influence of the structure on thermal conduction. To understand the formation of preferential orientation of fiber inclusions, polymer flow in the nozzle is experimentally studied. Section 2 describes materials and methods, Section 3 presents the Maxwell Garnett theory for the effective thermal conductivity/diffusivity in dispersed media. Section 4 presents and analyses the results.

## 2. Materials and methods

Two kinds of filled polymers are studied. The first is commercial composite PRO TOTAL CF-5 based on thermoplastic polyurethane (TPU) with 5% volume fraction of carbon fibers produced by Filamentarno [22]. The second is home-made composite based on polyamide 12 (PA) filled with equiaxed $Al_2O_3$ particles with variable volume fraction. The materials are in the form of filaments 1.75 mm in diameter ready for 3D printing. Smirnov *et al.* [23] gave the details for $Al_2O_3$/PA filament fabrication method. Figure 1 shows cross sections of the filaments obtained by light microscopy. Yellow needles in Fig. 1a are carbon fibers preferentially oriented parallel to the filament axis. The fibers are visible as dark dots in Fig. 1b. Thus, Figs. 1a and 1b indicate parallel orientation of carbon fibers. Such orientation is due to filament fabrication by extrusion. Large dark spots in Figs 1a and 1b are pores around the filament axis formed in the filament fabrication process. Figures 1c and 1d indicate no pores and no anisotropy in the microstructure. The horizontal line on the top of Fig. 1c is the filament boundary. The isotropic microscructure is reasonable in the case of equiaxed $Al_2O_3$ inclusions.

Printed samples of composite materials are obtained by commercial FFF printer Black Widow (Tevo 3D, China). Table 1 shows the parameters of printing. Firstly, cuboids with 1 cm width, 1 cm height and few centimeters in length are printed. As shown in Fig. 2, material is extruded along sequentially inscribed rectangles in every layer. Thus, in cross sections near the right and left bases of the printed bar, Section A in Fig. 2, extrusion direction is parallel to the cross section, while in cross sections near the middle of the bar, Section B, extrusion direction is normal to the cross section. Samples cut near the bases and in the middle of the bar are used to study anisotropy of the printed material. To understand polymer flow and orientation of fibers inside the printing nozzle, the nozzle used for printing from carbon fiber/TPU composite is separated from the printing head and cut along the axis. The samples are studied by metallography methods. They are polished and observed with light microscope Olympus BX51M (Olympus Corporation, Tokyo, Japan) and scanning electron microscope (SEM) VEGA 3 LMH (Tescan, Brno, Czech Republic).

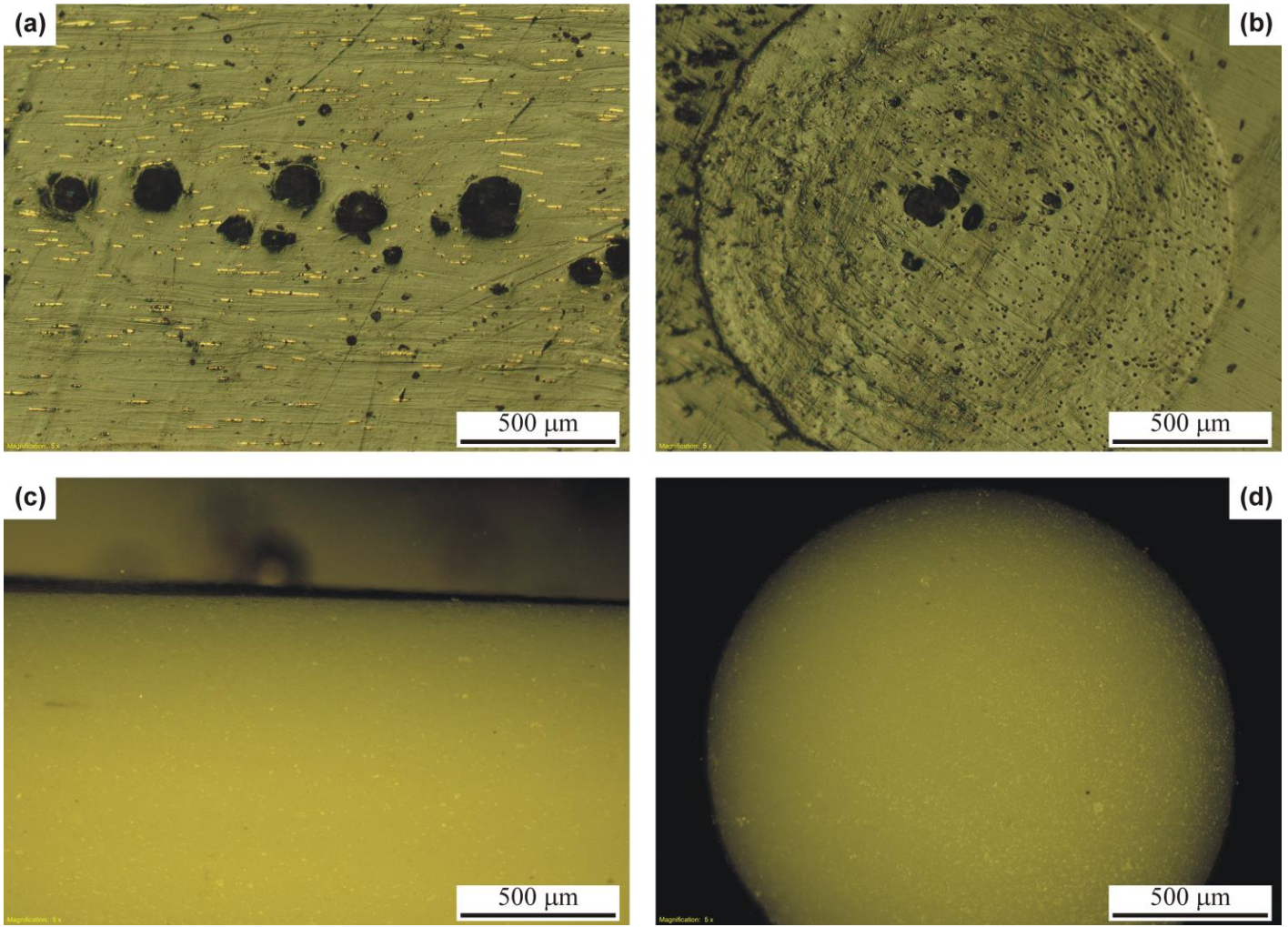

**Fig. 1.** Filaments for FFF 3D printing from the following composites: 5%vol. Carbon/TPU (a,b) and 10%vol. Al$_2$O$_3$/PA (c,d). Longitudinal (a,c) and transversal (b,d) cross sections.

**Table 1.** Printing parameters

| Material | Carbon fiber/TPU | Al$_2$O$_3$/PA |
|---|---|---|
| Hot end temperature, °C | 240 | 255 |
| Base plate temperature, °C | 100 | 110 |
| Printing speed, mm/s | 10 | 10 |
| Feeding rate | 110% | 110% |
| Nozzle diameter, mm | 0.4 | 0.6 |
| Raster width, mm | 0.4 | 0.6 |
| Layer height, mm | 0.2 | 0.2 |

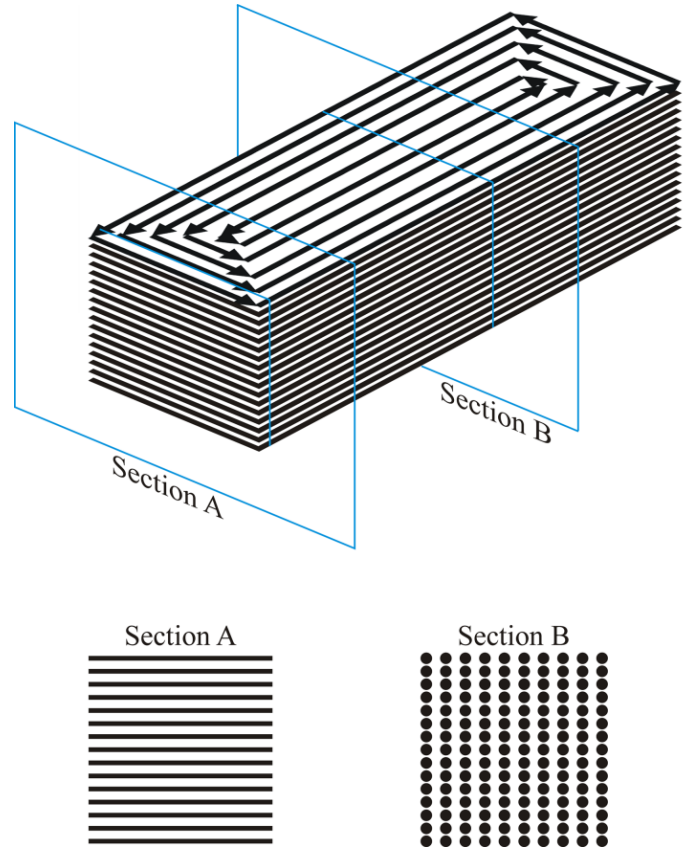

**Fig. 2.** Extrusion strategy for printing.

Thermal diffusivity of the printed composite materials is measured by laser flash method using LFA 467 HT HyperFlash setup (NETZSCH Inc., Germany) at 25˚C, 50˚C, 75˚C and 100˚C temperatures. Samples for thermal diffusivity are 10×10×1 mm plates. They are cut from the printed bars as shown in Fig. 2. The plates are covered with a thin graphite layer to ensure absorption of laser radiation. Thermal diffusivity is measured in the direction normal to the plate. Thus, one obtains thermal diffusivity perpendicular to the extrusion direction for the plates taken near the bases of the bar and parallel to the extrusion direction for the plates taken in the middle of the bar.

## 3. Maxwell Garnett approximation

The Maxwell Garnett theory estimates thermal conductivity/diffusivity of the studied composite materials with equiaxed and fiber inclusions. Below, the theory is applied to thermal diffusivity because this physical quantity is directly measured in the present work. The same is applicable to thermal conductivity because thermal conductivity and thermal diffusivity are related by the factor of specific heat.

Consider a dispersed medium consisting of matrix phase m and inclusions i. Let the inclusions be the same uniaxial ellipsoids. The orientation of the inclusions is the same too. Figure 3 shows the Cartesian coordinates chosen so that axis (OZ) is the rotation axis of the inclusions. The uniaxial ellipsoid is specified by two semi-axes, $r_\parallel$ parallel and $r_\perp$ perpendicular to axis (OZ), see Fig. 3. Suppose that the materials of matrix and inclusion phases are isotropic, with thermal diffusivities $\alpha_m$ and $\alpha_i$, respectively. Due to the regular orientation of the inclusions, the effective thermal diffusivity of the dispersed medium is, generally, anisotropic. Owing the symmetry, the principal axes of the thermal diffusivity tensor are the coordinate axes and the tensor of the effective thermal diffusivity is characterized by two principal values, the effective thermal diffusivity parallel to axis (OZ) $\alpha_\parallel$ and that perpendicular to axis (OZ) $\alpha_\perp$.

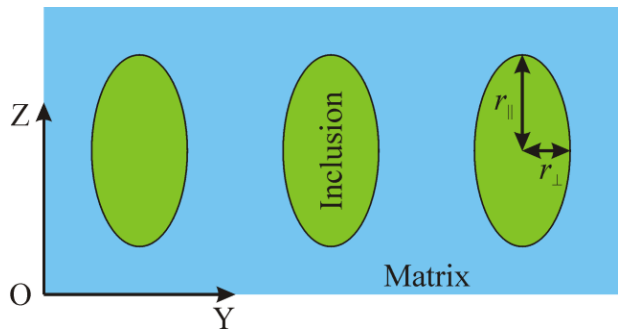

**Fig. 3.** Dispersed medium containing matrix (blue) and inclusions (green) with semi-axes $r_\parallel$ and $r_\perp$. Coordinate axes are (OY) and (OZ).

In the Maxwell Garnett approximation, the principal values of the effective thermal diffusivity tensor are [24]

$$\frac{\alpha_\gamma}{\alpha_m} = \frac{1 + f_i(1-n_\gamma)A_\gamma}{1 - f_i n_\gamma A_\gamma}, \quad A_\gamma = \frac{\alpha_i - \alpha_m}{n_\gamma \alpha_i + (1-n_\gamma)\alpha_m}, \tag{1}$$

with index $\gamma$ taking values $\parallel$ or $\perp$, where $f_i$ is the volume fraction of inclusions and $n_\gamma$ is the principal values of the depolarization tensor [24],

$$n_\parallel = \begin{cases} \frac{1-e^2}{2e^3}\left(\ln\frac{1+e}{1-e} - 2e\right), & e = \sqrt{1 - 1/a^2}, \quad \text{if } a > 1 \\ \frac{1}{3}, & \text{if } a = 1, \quad n_\perp = \frac{1-n_\parallel}{2}. \\ \frac{1+e^2}{e^3}(e - \arctan e), & e = \sqrt{1/a^2 - 1}, \quad \text{if } a < 1 \end{cases} \tag{2}$$

where $a = r_\parallel / r_\perp$ is the inclusion aspect ratio. In the case of spherical inclusions with $a = 1$, Eq. (1) reduces to

$$\frac{\alpha_\parallel}{\alpha_m} = \frac{\alpha_\perp}{\alpha_m} = \frac{\alpha_i + 2\alpha_m + 2f_i(\alpha_i - \alpha_m)}{\alpha_i + 2\alpha_m - f_i(\alpha_i - \alpha_m)}. \tag{3}$$

In the case of extremely prolate inclusions with $a \to \infty$, Eq. (1) becomes

$$\alpha_\parallel = f_i \alpha_i + (1-f_i)\alpha_m, \quad \frac{\alpha_\perp}{\alpha_m} = \frac{\alpha_i + \alpha_m + f_i(\alpha_i - \alpha_m)}{\alpha_i + \alpha_m - f_i(\alpha_i - \alpha_m)}. \tag{4}$$

The first Eq. (4) corresponds to parallel connection of the matrix and inclusion phases. If the inclusions are much more diffusive than the matrix, $\alpha_i \gg \alpha_m$, the effective thermal diffusivity becomes independent of the inclusion diffusivity in the case of spherical inclusions and extremely prolate inclusions in the transversal direction as follows from Eq. (3) and the last Eq. (4), respectively:

$$\frac{\alpha_\parallel}{\alpha_m} = \frac{\alpha_\perp}{\alpha_m} = \frac{1 + 2f_i}{1 - f_i}, \quad a = 1, \tag{5}$$

$$\frac{\alpha_\perp}{\alpha_m} = \frac{1 + f_i}{1 - f_i}, \quad a \to \infty. \tag{6}$$

At low fraction of inclusions $f_i \ll 1$, Eq. (5) grows with $f_i$ approximately as $1 + 3f_i$ while Eq. (6) grows as $1 + 2f_i$. This means that spherical inclusions are more effective to increase the thermal diffusivity than prolate inclusions in the transversal direction. The first Eq. (4) indicates that the effective thermal diffusivity in the longitudinal direction of the extremely prolate inclusions is proportional to their thermal diffusivity $\alpha_i$ and becomes much greater than the thermal diffusivity in the transversal direction with increasing $\alpha_i$.

Another important case is a high but finite aspect ratio $a \gg 1$ of prolate inclusions resulting in a low principal value of the depolarization tensor $n_\parallel$ according to Eq. (2). If the product of $n_\parallel$ and the inclusion thermal diffusivity is still much greater than the matrix thermal diffusivity,

$$n_\parallel \alpha_i \gg \alpha_m, \tag{7}$$

the effective thermal diffusivity in parallel direction,

$$\frac{\alpha_\parallel}{\alpha_m} = \frac{1 + f_i / n_\parallel}{1 - f_i}, \ a \gg 1, \tag{8}$$

becomes independent of the inclusion diffusivity. Taking into account a low value of $n_\parallel$, one concludes that ratio (8) steeply increases with the volume fraction of inclusions $f_i$. Figure 4 summarizes the relations between the effective thermal diffusivities of media with highly diffusive inclusions.

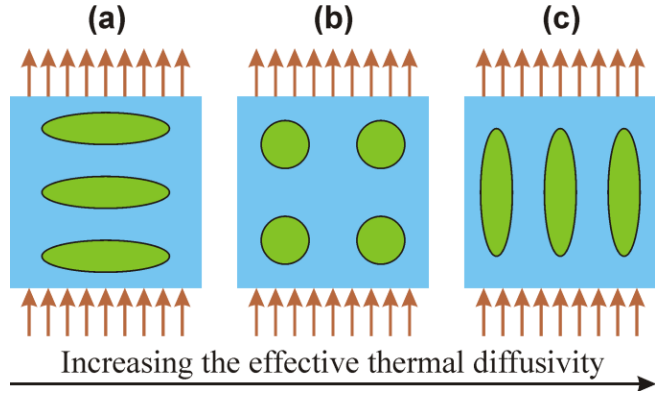

**Fig. 4.** Effective thermal diffusivity of the dispersed media with highly diffusive inclusions. Red arrows indicate the direction of heat flow. (a) Prolate inclusions perpendicular to heat flow. (b) Spherical inclusions. (c) Prolate inclusions parallel to heat flow.

## 4. Results and discussion

### 4.1. Polymer-based composite flow in the hot end

Carbon fibers visualize streamlines of TPU matrix in Fig. 5. The filament of 5%vol. Carbon/TPU passes from the cold end through a Teflon tube, see Fig. 5a. The Teflon tube is inside a cylindrical channel in a steel throat with a thin neck forming a thermal barrier pointed with an arrow in Fig. 5. After the Teflon tube, polymer flows into a brass nozzle screwed into an aluminum-alloy heater. Figure 5b shows that carbon fibers are approximately parallel to the axis above the thermal barrier while they become to deviate from the axial direction and form wavy streamlines below the barrier. The waviness considerably decreases when polymer enters the nozzle, see Fig. 5c. Fibers deviate again from the axis in the conical part of the nozzle on the bottom of Fig. 5d where they visualize converging streamlines. Finally, fibers are preferentially oriented parallel to the axis in the extrusion orifice.

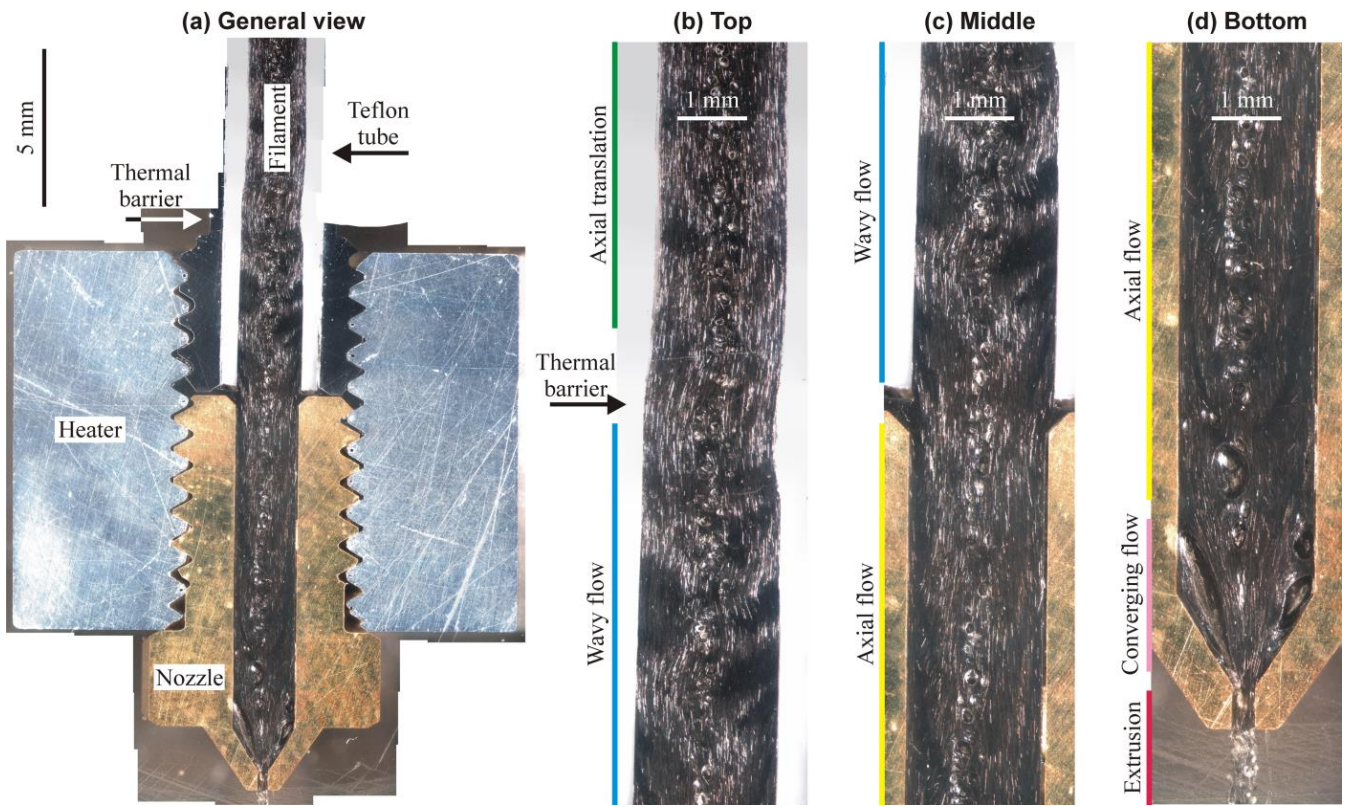

**Fig. 5.** Inner flow of composite Carbon/PA: Axial cross section of the hot end (a); Enlarged composite in the top (b), middle (c), and bottom (d).

The polymer-based composite should be in the solid state above the thermal barrier. Therefore, fibers are parallel like in the filament. One actually observes translation of the solid filament in axial direction as indicated on the top of Fig. 5b. Below the thermal barrier, the filament softens and becomes to flow. The waviness of the flow in the Teflon tube below the thermal barrier can be explained by the fact that the filament diameter is lower than the channel diameter. The softened polymer expands to fill the whole width of the channel. Such expansion means deviation of the flow direction from the axis. If the lateral expansion is not steady, this may lead to periodic variation of the expansion direction explaining the wavy disorientation of fibers. Another reasonable explanation is unsteady interaction between the plastic flow and the soft Teflon tube [25]. The distortions of TPU/Teflon interface visible in Fig. 5b are consistent with the mechanism of unsteady flow/wall interaction but cannot exclude the mechanism of unsteady lateral expansion.

The polymer flow stabilizes inside the nozzle and becomes axial, see Figs. 5c and 5d. Indeed, the walls of the brass nozzle are much more rigid, so that flow/wall interactions should not be important. The flow converges in the converging part of the nozzle and becomes axial in the extrusion orifice, see Fig. 5d. The orientation of fibers indicates a steady laminar flow in the nozzle. However, gas bubbles arising from pores in the filament can locally perturb the flow. Figure 5d indicates that the bubble size increases downstream. The increase can be explained by coalescence. When bubbles coalesce, their total volume should conserve. However, the volume fraction of gas phase visually increases downstream in Fig. 5d. Thus, the bubbles can swell. One can expect that polymer temperature increases downstream. This changes chemical equilibrium in the system liquid/gas to the greater gas volume. In other words, volatile components evaporate when temperature increases. This explains the swelling of gas bubbles. The bubbles can perturb polymer extrusion and thus lead to formation of defects in 3D printing. Therefore, the commonly advised filament drying [1] helps to remove volatile components and thus improve printing quality.

*4.2. Microstructure of printed composites*

Figure 6 shows light-microscopy images of the printed Carbon/TPU composite in the cross sections parallel (a) and normal (b) to the extrusion direction. Horizontal bands of darker contrast correspond to boundaries between layers in Fig. 6a. Similar bands surround extrusion beads in Fig. 6b. The orientation of carbon fibers in the extrusion bead approximately corresponds to their orientation in the filament. Therefore, one can observe more fibers oriented parallel to the image plane (yellow dashes) in Fig. 6a and more fibers intersecting the image plane (yellow dots) in Fig. 6b. Thus, the printed block inherits preferential orientation of fibers observed in the filament. In the filament, the alignment direction of fibers is the filament axis. In the printed block, the alignment direction is the extrusion direction.

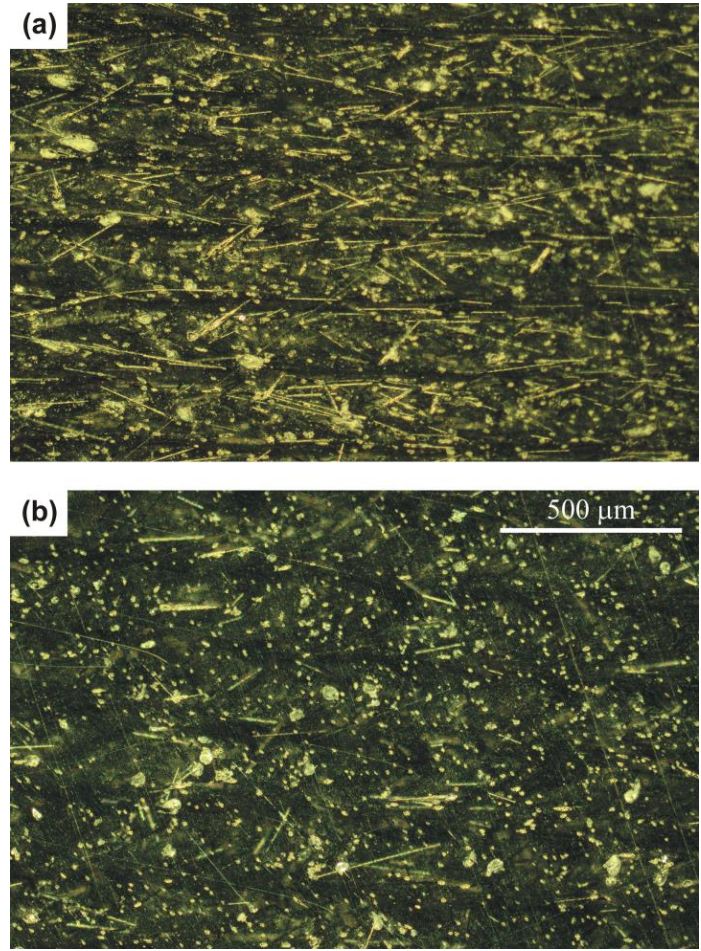

**Fig. 6.** Microstructure of printed composite 5%vol. Carbon/TPU: cross sections parallel (a) and normal (b) to the extrusion direction.

Figure 7 compares the microstructures of the filament (a,b) and printed block (c,d). Filament axis (1) and extrusion direction (2) are shown as a dashed-line interval when parallel to the image plane and a cross in a circle when normal to the image plane. Yellow dashes (3) observed in Figs. 7a and 7c are cross sections of fibers approximately parallel to the filament axis (1 in Fig. 7a) and extrusion direction (2 in Fig. 7c). The carbon fibers are few microns in diameter and can be greater than 100 microns in length. In Fig. 7c, there are less dashes and they are more disordered in orientation. Besides, there are fibers at small angle to the image plain (4). This means that fibers in the printed block are more disoriented relative to the extrusion direction and fibers in the filament are less disoriented relative to the filament axis. The dots (5) observed in Figs. 7b and 7d are fibers intersecting the image plane at a great angle, which are approximately parallel to the filament axis (1 in Fig. 7b) and extrusion direction (2 in Fig. 7d). In the printed block, Fig. 7d, there are several diffuse dashes corresponding to fibers considerably deviating from the extrusion direction. This indicates more disorientation in the printed material. In summary, disorientation of fibers relative to the alignment direction increases after printing.

Figure 8 shows light-microscopy images of the printed $Al_2O_3$/PA composite in the cross sections parallel (a) and normal (b) to the extrusion direction. In Fig. 8a, one can observe horizontal lines corresponding to the boundaries between layers. In Fig. 8b, there is a lattice of triangles. Observing the two projections, one can state that there is a lattice of triangle-based cylindrical pores parallel to the extrusion direction. The pores are formed between extrusion beads. Figure 9 shows higher-magnification SEM images of microstructure in the domains between the pores. There are irregular-shape $Al_2O_3$ particles dispersed in a polymer matrix. The most of the particles are approximately equiaxed. The size of the observed particles varies from few microns to 40-50 microns. The particles are uniformly distributed the matrix. No agglomeration or segregation are observed. According to the Delesse principle [26], the surface fraction occupied by particle cross sections in micrographs (a)-(c) corresponds to the nominal volume fraction of $Al_2O_3$ inclusions indicated in the caption of Fig. 9.

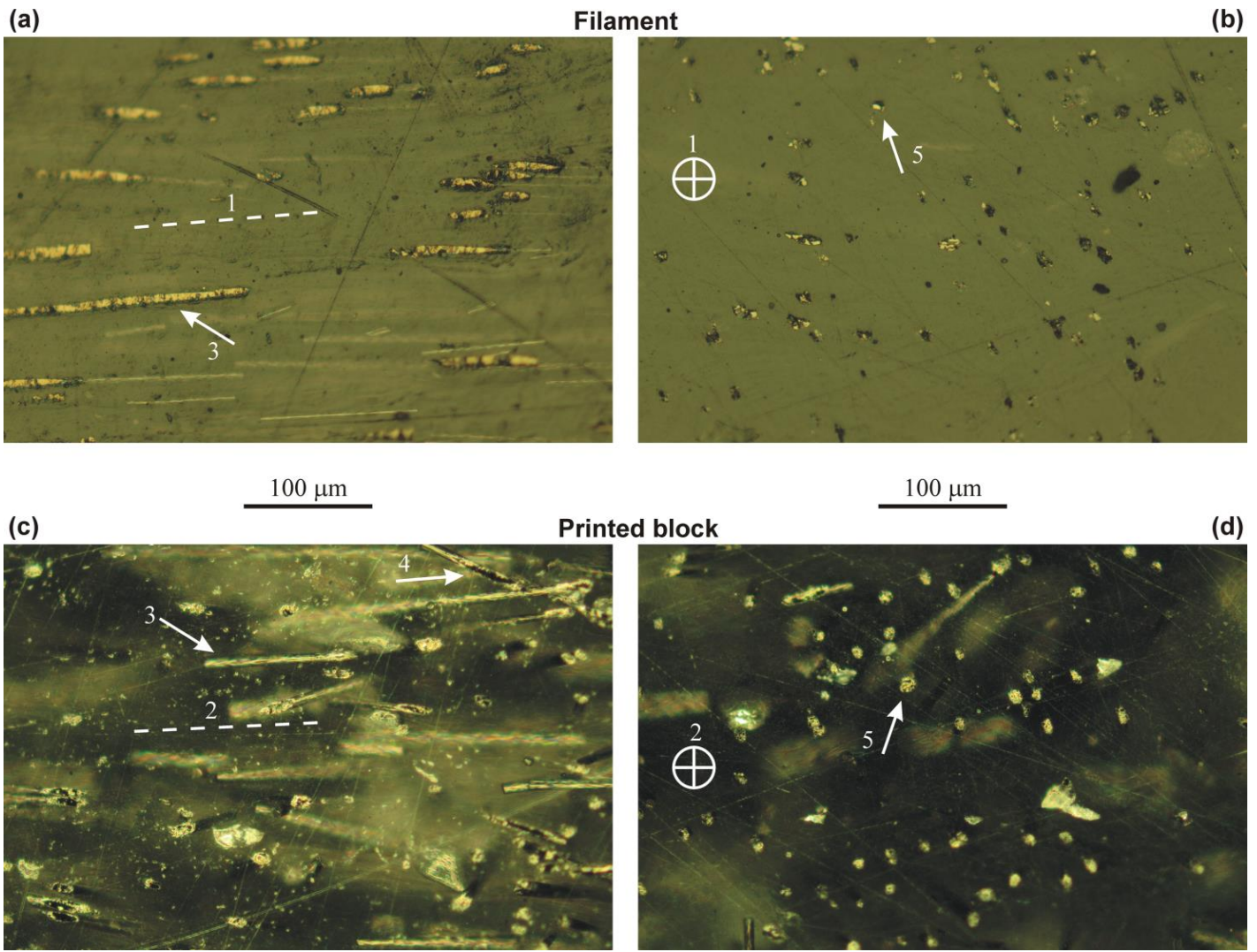

**Fig. 7.** Cross sections of 5%vol. Carbon/TPU filament (a,b) compared to printed block (c,d): parallel (a) and normal (b) to the filament axis (1) and parallel (c) and normal (d) to the extrusion direction (2). Carbon fibers approximately parallel (3), intersecting at a small angle (4), and approximately normal (5) to the image plane.

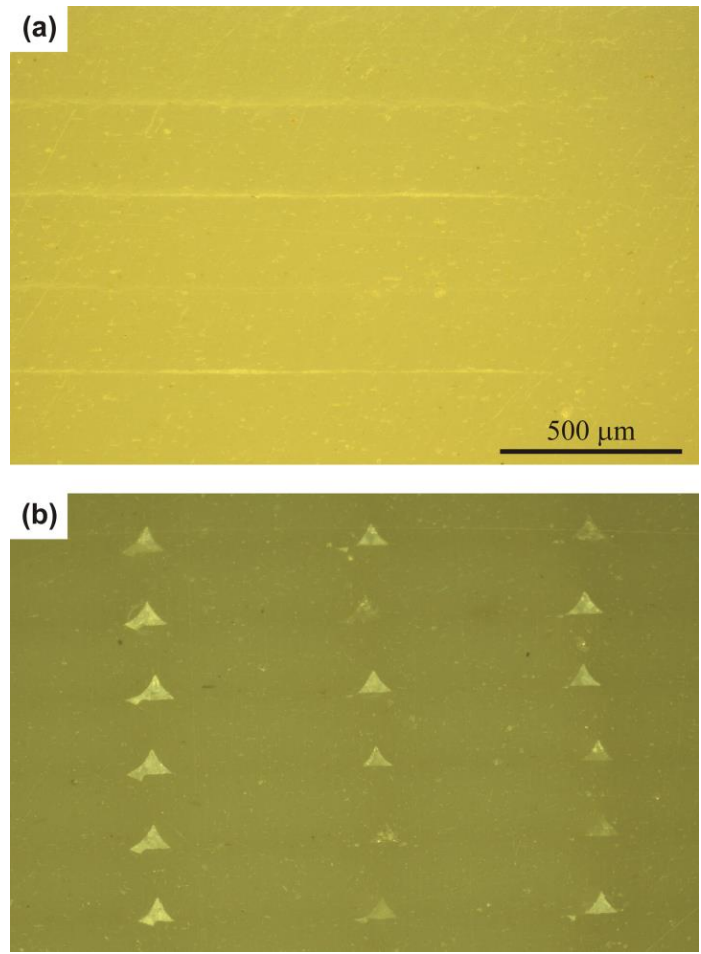

**Fig. 8.** Microstructure of printed composite 10%vol. $Al_2O_3$/PA: cross sections parallel (a) and normal (b) to the extrusion direction.

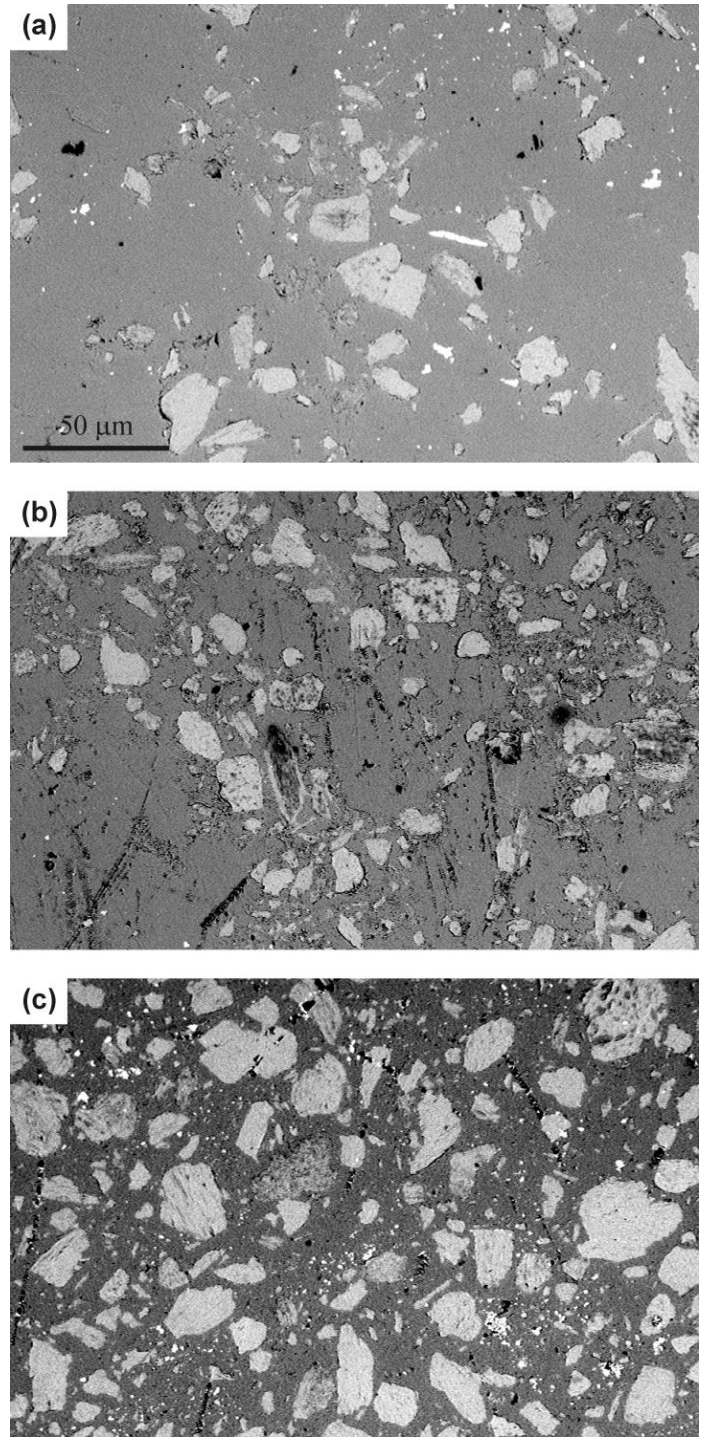

**Fig. 9.** Cross sections of printed composite Al$_2$O$_3$/PA at the following volume fractions of Al$_2$O$_3$: 10% (a), 30% (b) and 50% (c)

*4.3. Thermal diffusivity of printed composites*

Four samples are cut from a printed block of 5%vol. Carbon fiber/TPU composite: Samples 1 and 2 for measuring thermal diffusivity parallel to the extrusion direction and Samples 3 and 4 for thermal diffusivity perpendicular to the extrusion direction. After stabilizing temperature at the values indicated in Table2, thermal diffusivity is measured by the laser flash setup after three laser shots. Table 2 lists the measured thermal diffusivity values as the mean of the three shots. The uncertainty of around 0.001 mm$^2$/s is estimated as the mean square deviation of the three shots. Table 2 shows that there is a small but significant and systematic difference in thermal diffusivity between Samples 1 and 2 as well as between Samples 3 and 4. This difference reflects local variations in material microstructure.

**Table 2.** Thermal diffusivity of printed composite 5%vol. Carbon fiber/TPU

| Temperature, $T$ (°C) | Thermal diffusivity, (mm$^2$/s) | | | |
|---|---|---|---|---|
| | Heat flow parallel to extrusion direction, $\alpha_\parallel$ | | Heat flow perpendicular to extrusion direction, $\alpha_\perp$ | |
| | Sample 1 | Sample 2 | Sample 3 | Sample 4 |
| 25 | 0.230 | 0.221 | 0.146 | 0.141 |
| 50 | 0.216 | 0.208 | 0.134 | 0.13 |
| 75 | 0.206 | 0.197 | 0.126 | 0.122 |
| 100 | 0.197 | 0.189 | 0.118 | 0.115 |

Figure 10 compares the measured values of thermal diffusivity with theoretical estimates given in Section 3. The Maxwell Garnett theory indicates that the effective thermal diffusivity of composite depends both on the matrix and inclusion thermal diffusivities. The thermal diffusivity of TPU matrix is around 0.11 mm$^2$/s at the room temperature [27]. To the best of our knowledge, thermal diffusivity of a carbon fiber has not been directly measured because of the small size. The atomic structure of carbon fibers is similar to that of graphite with graphene sheets parallel to the fiber axis [28]. Therefore, thermal diffusivity of a fiber along the axis is estimated as that of graphite parallel to the sheets while thermal diffusivity of a fiber perpendicular to the axis is estimated as that of graphite perpendicular to the sheets. In highly oriented pyrolytic graphite, the values of thermal diffusivity are of the order of 1000 mm$^2$/s and 5 mm$^2$/s, respectively, at the room temperature [29]. The both values are much greater than the thermal diffusivity of TPU.

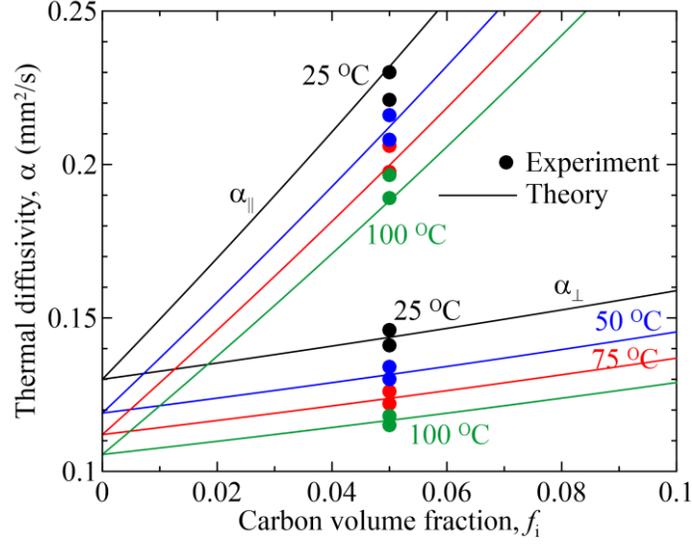

**Fig. 10.** Effective thermal diffusivity of printed composite Carbon fiber/TPU: experimental points and theoretical curves for the extrusion direction parallel (upper branches) and perpendicular (lower branches) to heat flow. The values of temperature are indicated near the curves.

Firstly, the best theoretical fit is found for the thermal diffusivity $\alpha_\perp$ in the direction perpendicular to the extrusion direction. One assumes that carbon fibers are much more thermally diffusive than the TPU matrix. Therefore, Eq. (6) is applicable. This equation contains the unknown value of TPU matrix thermal diffusivity $\alpha_m$. The $\alpha_m$ values are chosen to obtain the closest fit to the present experimental data by Eq. (6). Table 3 lists these values for the considered temperatures $T$. They reasonably agree with the independently measured TPU thermal diffusivity [27].

**Table 3.** Accepted thermal diffusivity values for TPU matrix $\alpha_m$ to estimate the effective thermal diffusivity of Carbon/TPU composite perpendicular to the extrusion direction by Eq. (6)

| Temperature, $T$ (°C) | 25 | 50 | 75 | 100 |
|---|---|---|---|---|
| Matrix, $\alpha_m$ (mm²/s) | 0.130 | 0.119 | 0.112 | 0.1055 |

Secondly, the best theoretical fit is found for the thermal diffusivity $\alpha_\parallel$ in the direction parallel to the extrusion direction. This is the direction of carbon fibers alignment. Obviously, general Maxwell Garnett formula given by Eqs. (1) and (2) simplifies due to the high inclusion-to-matrix thermal diffusivity ratio. However, there are two quite different cases given by Eqs. (4) and (8). The choice is made by Eq. (7). Figure 7a shows that the aspect ratio of the carbon fibers is of the order of $a = 10 \ldots 100$. According to Eq. (2), the corresponding values of depolarization tensor principal value $n_\parallel$ are in the range from $4\cdot10^{-4}$ to $2\cdot10^{-2}$. Therefore, the left-hand side of Eq. (7) is of the order of 0.4 to 20 mm²/s, where thermal diffusivity of fibers is taken along the axis. This value is still much greater than TPU thermal diffusivity. Thus, inequality (7) is valid and Eq. (8) should be used to calculate thermal diffusivity parallel to the extrusion direction $\alpha_\parallel$.

Equation (8) is also independent of inclusion thermal diffusivity $\alpha_i$. The parameters of this equation are matrix thermal diffusivity $\alpha_m$ and depolarization tensor principal value $n_\parallel$. The former has been found and listed in Table 3 depending of temperature. The depolarization tensor is a structure parameter and should not depend on temperature. Table 4 shows the found value of $n_\parallel$ to obtain the best fit of the experimental data at the studied temperatures. The corresponding upper branches of the curves in Fig. 10 do agree with the present experimental data (points). Table 4 also shows the corresponding aspect ratio of fibers a calculated from the found $n_\parallel$ value by Eq. (2). The aspect ratio estimated from the thermal diffusivity data seems to underestimate the aspect ratio observed in the microstructure of Fig. 7a. The discrepancy can be explained by disorientation of fibers in the printed sample, see Fig. 7c. The disorientation reduces the effective thermal diffusivity in the extrusion direction. Reducing thermal diffusivity results in decreasing *a* according to Eq. (8). Thus, the value of *a* shown in Table 4 is not the actual aspect ratio but rather its effective value to take into account fibers disorientation for the use in Eq. (8) derived in the ideal conditions of parallel fibers.

**Table 4.** Accepted structural parameters to estimate the effective thermal diffusivity of Carbon/TPU composite parallel to the extrusion direction by Eq. (8)

| Parameter | Value |
|---|---|
| Principal value of the depolarization tensor parallel to the extrusion axis, $n_\parallel$ | 0.072 |
| Inclusion aspect ratio, *a* | 4.1 |

Preliminary experiments with printed $Al_2O_3$/PA composites have indicated that their thermal diffusivity parallel and perpendicular to the extrusion direction is essentially the same. The observed difference is around the measurement error. Thus, no anisotropy in thermal diffusivity is detected. This is reasonable because the only anisotropic element of microstructure is triangle-based cylindric pores shown in Fig. 8, which occupy a small volume fraction. Taking into account isotropy of $Al_2O_3$/PA composites, the orientation of samples relative the extrusion direction is not significant for measuring thermal diffusivity. Five samples with $Al_2O_3$ volume fraction from 10 to 50% are cut from printed blocks. The three-shot laser-flash measurement of thermal diffusivity is applied to $Al_2O_3$/PA similar to the experimental procedure for Carbon/TPU. Table 5 lists the measured thermal diffusivity values as the mean of the three shots. The mean square deviation is around 0.001 mm$^2$/s.

**Table 5.** Thermal diffusivity $\alpha$ of printed composite $Al_2O_3$/PA

| Volume fraction of $Al_2O_3$, $f_i$ | Temperature, $T$ (°C) | Thermal diffusivity, $\alpha$ (mm$^2$/s) |
|---|---|---|
| 10% | 25 | 0.206 |
|  | 50 | 0.179 |
|  | 75 | 0.158 |
|  | 100 | 0.142 |
| 20% | 25 | 0.270 |
|  | 50 | 0.236 |
|  | 75 | 0.212 |
|  | 100 | 0.192 |
| 30% | 25 | 0.280 |
|  | 50 | 0.258 |
|  | 75 | 0.236 |
|  | 100 | 0.218 |
| 40% | 25 | 0.324 |
|  | 50 | 0.294 |
|  | 75 | 0.268 |
|  | 100 | 0.246 |
| 50% | 25 | 0.346 |
|  | 50 | 0.309 |
|  | 75 | 0.279 |
|  | 100 | 0.255 |

Figure 11 compares the obtained experimental data with the Maxwell Garnett theory. Thermal diffusivity of alumina is of the order of 10 mm$^2$/s at the room temperature [30] while thermal diffusivity of PA is around 0.14 mm$^2$/s [31]. Therefore, the theoretical estimation for equiaxed inclusions by Eq. (5) is used assuming that Al$_2$O$_3$ particles are much more thermally diffusive than PA matrix. In this regime, the composite thermal diffusivity becomes independent of inclusion thermal diffusivity $\alpha_i$. The unknown value of matrix thermal diffusivity $\alpha_m$ is chosen to fit the experimental value of effective thermal diffusivity $\alpha$ at Al$_2$O$_3$ volume fraction of 20%. Table 6 lists the obtained values of $\alpha_m$. They do agree with independent measurements [31].

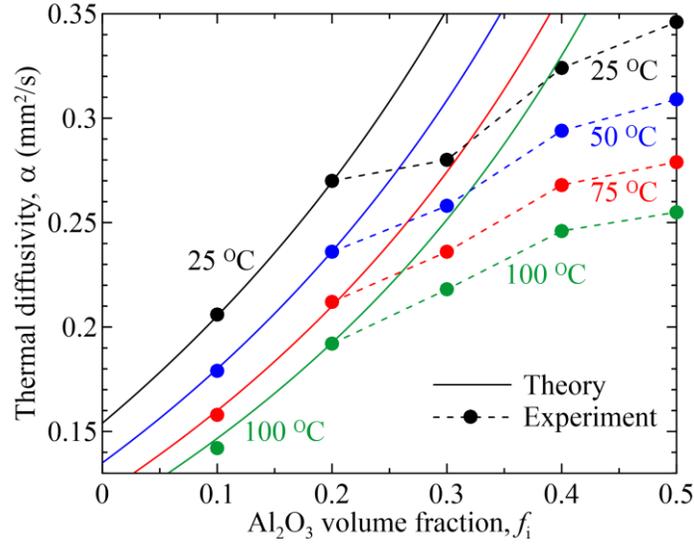

**Fig. 11.** Thermal diffusivity of Al$_2$O$_3$/PA composites as function of Al$_2$O$_3$ volume fraction: Experiment (points connected by dashed lines) and Maxwell Garnett theory (curves). The values of temperature are indicated near the curves.

**Table 6.** Accepted thermal diffusivity values for PA matrix $\alpha_m$ to estimate the effective thermal diffusivity of Al$_2$O$_3$/PA composite by Eq. (5)

| Temperature, $T$ (°C) | 25 | 50 | 75 | 100 |
|---|---|---|---|---|
| Matrix, $\alpha_m$ (mm$^2$/s) | 0.154 | 0.135 | 0.12 | 0.11 |

Figure 11 shows that the effective thermal diffusivity increases with the volume fraction of inclusions. This is reasonable because inclusions are much more thermally diffusive than matrix. The theoretical curves perfectly follow the experimental data for the volume fraction of inclusions less or equal to 20% and overestimate the experimental data at the volume fraction greater or equal to 30%. The principal assumption of the Maxwell Garnett theory is the low volume fraction of inclusions when neighbor inclusions cannot influence each other [32]. The microstructures of Fig. 9 indicate that the distance between neighbor inclusions considerably decreases with the volume fraction and becomes much less than the size of an inclusion in Fig. 9c at 50% of volume fraction. One can expect an intensive mutual influence of neighbor inclusions in these conditions. The influence can explain the observed deviation from the theory.

Figure 12 summarizes the obtained data on polymer matrix thermal diffusivity given by Tables 3 and 6 and compares them to literature data [27,31]. The present measurements (circles) generally agree with the reference experiments (stars), however a small systematic difference is observed. It is known that thermal diffusivity is sensitive to the phase composition, degree of crystallinity, and grain size of the material. The mentioned parameters may differ for the same polymer obtained in different conditions. This explains the small discrepancy observed. Indeed, the extruded material is cooled in thin layers at FFF. This gives an increased cooling rate and may result in a considerable deviation from the equilibrium phase composition.

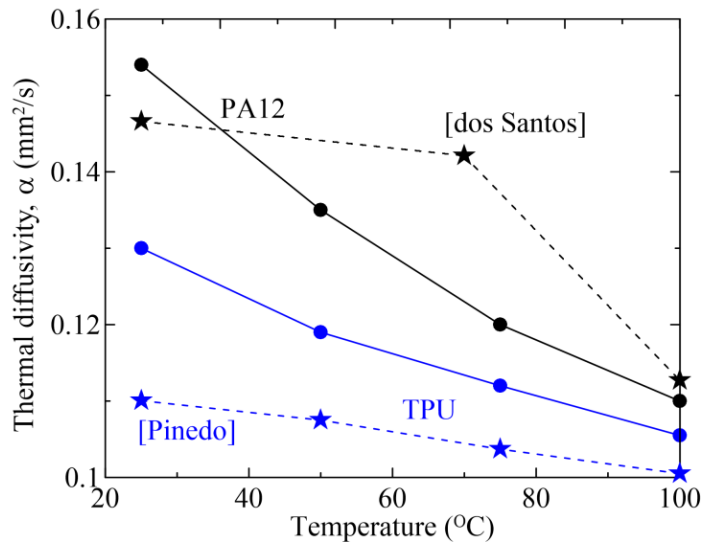

**Fig. 12.** Thermal diffusivity of PA12 (black) and TPU (blue) matrices derived from the present experiments (circles) compared to the literature data (stars) by dos Santos *et al.* [31] and Pineda *et al.* [27].

For the present FFF process with 10%vol. Al$_2$O$_3$/PA composite, one can estimate cooling time $\Delta t$ from the layer height $h = 0.2$ mm (see Table 1) and thermal diffusivity $\alpha = 0.206$ mm$^2$/s (see Table 5) by dimensional analysis as

$$\Delta t = \frac{h^2}{\alpha} \approx 0.2 \text{ s}. \tag{9}$$

The hot end temperature is $T_h = 255\ ^\text{O}\text{C}$ (see Table 1). Therefore, the mean cooling rate is

$$\frac{T_h}{\Delta t} \approx 1300 \text{ K/s}. \tag{10}$$

Zhang *et al.* [33] reported the critical cooling rate of 300 K/s for Polyamide 12. In the present conditions, the cooling rate is considerably greater than the critical one, which emphasizes the importance of crystallization kinetics in formation of polymer matrix microstructure. Indeed, it is probable a high deviation from phase equilibrium and formation of an amorphous structure, which can influence thermal properties.

Figures 10 and 11 indicate that the thermal diffusivity of all the studied polymer-based composites decreases with temperature. Such a temperature dependence seems to be reasonable because the Maxwell Garnett model predicts that the composite effective thermal diffusivity is proportional to the thermal diffusivity of the matrix polymer phase, see Eqs. (5), (6), and (8), which decreases with temperature according to Fig. 12. The observed decrease with temperature is typical for crystalline dielectrics where the principal carriers of thermal energy are phonons. Phonon-phonon scattering intensifies with temperature and suppresses thermal conduction [34].

Einstein's model of coupled harmonic oscillators is generally applicable to amorphous phases [35]. Thermal conduction in amorphous solids can increase in some temperature intervals and decrease in others [35]. A moderate increase of thermal conduction with temperature is typical for amorphous polymers while a decrease can be observed in partly crystalline polymers [36]. In summary, the observed temperature dependence is typical for crystalline or partly crystalline polymers. A deeper understanding of thermal conduction in FFF-printed polymer-based composites may require further studies of the formed structure of polymer matrix.

## 5. Conclusion

Fibers in a polymer-matrix composite visualize polymer flow in the hot end of the FFF printer. One can distinguish wavy flow in the throat downstream the thermal barrier, laminar axial flow in the cylindrical part of the nozzle, laminar converging flow in the conical part of the nozzle, and laminar axial flow in the extrusion orifice.

In the nozzle, fibers are generally oriented parallel to the flow direction while gas bubbles may locally disturb flow and disorient fibers. Gas bubbles grow in the nozzle, probably, due to the increase of temperature downstream, which leads to evaporation of volatile components.

Printed composite Carbon fiber/TPU inherits preferential orientation of fibers observed in the filament. In the filament, the alignment direction of fibers is the filament axis. In the printed block, the alignment direction is the extrusion direction. Disorientation of fibers relative to the alignment direction increases after printing. The carbon fibers are few microns in diameter and can be greater than 100 microns in length.

In the printed $Al_2O_3$/PA composite, there is a lattice of triangle-based cylindrical pores parallel to the extrusion direction. The pores are formed between extrusion beads. $Al_2O_3$ particles are of irregular shape and approximately equiaxed. Their size varies from few microns to 40-50 microns. No segregation or agglomeration are observed.

Significant anisotropy of thermal diffusivity is observed in the printed samples of 5%vol. Carbon fiber/TPU composite. The anisotropy is due to alignment of highly thermally diffusive carbon fibers. The effective thermal diffusivity of the composite in the directions parallel and perpendicular to the extrusion direction can be estimated by the Maxwell Garnett theory. High fiber-to-matrix ratio of thermal diffusivity simplifies the theoretical approach. Equation (8) is applicable in the parallel direction while Eq. (6) is valid in the perpendicular direction. The effective thermal diffusivity is essentially independent of the fiber thermal diffusivity in the both directions.

The measured effective thermal diffusivity of printed $Al_2O_3$/PA composites is essentially isotropic and significantly increases with the volume fraction of inclusions $f_i$ because the inclusions are much more thermally diffusive than the matrix. At $f_i \leq 20\%$, the present experiments agree with the Maxwell Garnett theory, Eq. (5). At $f_i \geq 30\%$, Eq. (5) becomes to overestimate the experiments because the Maxwell Garnett theory neglects mutual influence of neighbor inclusions increasing with the volume fraction.

In the studied FFF-printed composites, the effective thermal diffusivity is proportional to the thermal diffusivity of polymer matrix, which is sensitive to its structure formed in non-equilibrium conditions of rapid quenching from liquid phase. Further studying of the polymer-matrix structure may be useful for a deeper understanding of thermal conduction in the FFF-printed composites.

**Authors' contribution**     Conceptualization AG, AS. Investigation EK, MG, NB, NWSP, RK. Supervision AS. Writing—original draft AG, RK. Writing—review and editing AG, AS.

**Funding**     This work was supported by Ministry of Science and Higher Education of the Russian Federation under project 0707-2020-0034.

**Acknowledgments**     The experiments were carried out using the equipment of the Centre of collective use of MSUT "STANKIN" (project No. 075-15-2021-695).

**Declarations**

**Ethical approval**     Not applicable.

**Consent to participate**     Not applicable.

**Consent to publication**     Not applicable.

**Conflict of interest**     The authors declare no competing interests.